\documentclass{appolb}
\usepackage{epsfig,amsmath}
\usepackage{lineno}

\newcommand\sfrac[2]{{\textstyle \frac{#1}{#2}}}

\begin{document}
\linenumbers
\title{Progress in the parametrisation of the Neutrino sector
\thanks{Presented at the XXXVII International Conference of Theoretical Physics,
MATTER TO THE DEEPEST: Recent Developments in Physics of Fundamental
Interactions, USTRON'13 }%
}
\author{Thomas Gajdosik, Andrius Juodagalvis, Darius Jur\v{c}iukonis, Tomas Sabonis
\address{Vilnius University, Universiteto 3, LT-01513, Vilnius, Lithuania}
}
\maketitle
\begin{abstract}
Adding gauge singlets to the original Standard Model allows an explanation for the observed smallness of the neutrino masses using the seesaw mechanism. Following our plans presented in the last conference of this series we present the results for the non-standard setting, when the number of the singlets is smaller than the number of the SM generations.  
\end{abstract}
\PACS{11.30.Rd, 13.15.+g, 14.60.St}
  
\section{Continuing Ustron'11 }
In \cite{Gajdosik:2011kq} we described our plans to parametrise the $n_{L}$-generations Standard Model equipped with $n_{R}$ additional gauge singlet fermions $N_{R}$ and $n_{H}\ge 1$ Higgs doublets $\phi_{k}$ \cite{Grimus:1989pu}. The Grimus-Lavoura ansatz \cite{Grimus:2000vj} gives the masses and mixing parameters in terms of the parameters of the Lagrangian 
\begin{equation}
\mathcal{L} = 
\mathcal{L}_{\mathrm{SM},{\nu}}  
- \tilde{\phi}^{\dagger}_{k} \bar{N}_{R} \mathrm{Y}_{\! \nu}^{k} \, L_{L} 
- \sfrac{1}{2} N_{R}^{\top} \mathbf{C}^{-1} M_{R} N_{R}+ h.c. 
\enspace ,
\label{nuYukawa}
\end{equation}
where $\tilde{\phi}_{k} = i\tau_{2}\phi^{*}_{k}$ is the $SU(2)$-conjugated Higgs doublet and $\mathrm{Y}_{\! \nu}^{k}$ is the $n_{R} \times n_{L}$ neutrino Yukawa matrix for the $k$-th Higgs doublet. 

Electroweak symmetry breaking triggered by the vacuum expectation values of the neutral Higgs fields 
$
( 0, \sfrac{1}{\sqrt{2}} v_{k} )^{\top} = \langle \phi_{k} \rangle_{0}^{\top}
$
gives an effective mass term to all Standard Model particles: the vector bosons, the Higgs bosons, and the charged fermions. It also generates the $n_{R}\times n_{L}$ dimensional Dirac mass term 
\begin{equation}
  M_{D}
= 
  \sum_{k=1}^{n_{H}} \sfrac{1}{\sqrt{2}} v_{k} \mathrm{Y}_{\! \nu}^{k}
\enspace ,
\label{MD}
\end{equation}
that enters the symmetric $(n_{L}+n_{R})\times(n_{L}+n_{R})$ neutrino mass matrix
\begin{equation} 
  M_{\nu}
=
  \left(\begin{array}{cc}
     M_{L} & M_{D}^{\top} \\
     M_{D} & M_{R}
  \end{array}\right)
\enspace ,
\label{Mnu}
\end{equation}
where $M_{R}$ is the Majorana mass matrix from eq.(\ref{nuYukawa}) and $M_{L} = 0$ at tree level, as such a term violates the $U(1)_{\mathrm{Y}} \times SU(2)_{\mathrm{weak}}$ gauge symmetry of the Standard Model.  

The most convenient diagonalization of the mass matrix $M_{\nu}$, eq.(\ref{Mnu}), for  arbitrary $n_{L}$ and $n_{R}$ is the Grimus-Lavoura ansatz~\cite{Grimus:2000vj}, as it reduces the $(n_{L}+n_{R})^{2}$ parameters of the unitary diagonalisation matrix to only $2 n_{L} n_{R}$ parameters in the complex matrix $B$ (see \cite{Grimus:2000vj}) that go into the ansatz.

\section{Reverse engineering the Grimus-Lavoura ansatz}
The Grimus-Lavoura ansatz determines the masses and mixing matrices of the physical particles from the parameters of the Lagrangian. Our idea was to determine the parameters of the Lagrangian from the masses and mixings. 

The Casas-Ibarra parametrization~\cite{Casas:2001sr}, used in~\cite{AristizabalSierra:2011mn}, does this determination for $n_{L}=n_{R}$ and solves the leading order seesaw~\cite{seesaw} equation  
\begin{equation} 
  M_{\ell} 
=
  - M_{D}^{\top} M_{h}^{-1} M_{D} 
\label{nr3seesaw}
\end{equation}
by the ansatz
\begin{equation} 
  M_{D} = i M_{h}^{1/2} \cdot O \cdot M_{\ell}^{1/2}
\label{casas-ibarra}
\end{equation}
with an arbitrary (complex) orthogonal matrix $O$. This is the most general parametrisation for the case $n_{L} = n_{R}$. Our investigation for the case $n_{L} > n_{R}$ showed, that it is always possible to reduce the problem of diagonalising the $(n_{L}+n_{R})\times(n_{L}+n_{R})$ dimensional $M_{\nu}$ to diagonalising an effective $2n_{R}$ dimensional $M_{\nu}^{\prime}$ using unitary matrices:
\begin{equation} 
  U^{\top}
  M_{\nu}
  U
= 
  U^{\top}
  \left(\begin{array}{cc}
         0 & M_{D}^{\top} \\
     M_{D} & M_{R}
  \end{array}\right)
  U
= 
  \left(\begin{array}{cc}
     0 & 0 \\
     0 & M_{\nu}^{\prime}
  \end{array}\right)
\enspace .
\label{Mnu1}
\end{equation}
That this is possible was argued before in~\cite{Grimus:1989pu}. We construct the explicit matrices for this reduction. The case $n_{L} = 3$ and $n_{R} = 1$ we presented in the conference \cite{slides}.

For the case and $n_{L} = 3$ and $n_{R} = 2$ we can define the unitary matrix $U$ as a product $U=U_{12} \cdot U_{13}$ with the unitary matrices defined as 
\begin{equation} 
  (U_{1n})_{jk} 
= \delta_{jk} 
- ( 1 - \sqrt{1-|s_{n}|^{2} } )( \delta_{j1}\delta_{k1} + \delta_{jn}\delta_{kn} ) 
+ s_{n} \delta_{j1}\delta_{kn} - s_{n}^{*} \delta_{jn}\delta_{k1}
\enspace ,
\label{U32}
\end{equation}
where angles and phases are given by
\begin{equation} 
  \frac{s_{2}^{*}}{\sqrt{1-|s_{2}|^{2} }}
= 
  \frac{(M_{\nu})_{14} (M_{\nu})_{35} - (M_{\nu})_{15} (M_{\nu})_{34}}
       {(M_{\nu})_{24} (M_{\nu})_{35} - (M_{\nu})_{25} (M_{\nu})_{34}}
\label{s2}
\end{equation}
and
\begin{equation} 
  \frac{s_{3}^{*}}{\sqrt{1-|s_{3}|^{2} }}
= - \sqrt{1-|s_{2}|^{2} }
  \frac{(M_{\nu})_{14} (M_{\nu})_{25} - (M_{\nu})_{15} (M_{\nu})_{24}}
       {(M_{\nu})_{24} (M_{\nu})_{35} - (M_{\nu})_{25} (M_{\nu})_{34}}
\enspace .
\label{s3}
\end{equation}

\section{Numerical evaluations}
The analytic analysis using angles and phases gives a deeper insight into the problem. But the rotation matrices defined by the angles can lead to numerical instabilities and more time consuming operations in the numerical calculations than using the parameters of the Lagrangian directly. Therefore we just use a direct parametrisation of the Yukawa matrices and not our result from the reverse engineering. 

\subsection{The case $n_{L}=3$ and $n_{R}=1$}
We parametrise the Yukawa coupling as 
\begin{equation} 
\mathrm{Y}_{\! \nu}^{k} = \frac{\sqrt{2}}{v} m_{D} \vec{a}_{k}^{\top}
\quad\mbox{with}\enspace
v^{2} = \sum_{k=1}^{n_{H}} v^{2}_{k} = \frac{2 m_{W}^{2}}{g^{2}}
\enspace ,
\label{defYukawa31}
\end{equation}
where $m_{D}$ is the singular value of $M_{D}$, and the vectors $\vec{a}_{k}$ describe the relative coupling stength of the Higgs doublets. At tree level we get the mass relations 
\begin{equation} 
  m_{D}^{2} = m_{\ell} m_{h}
\quad\mbox{and}\quad
  m_{R} = ( m_{h} - m_{\ell} ) \sim m_{h}
\enspace ,
\label{3+1treemasses}
\end{equation}
with $m_{\ell}$ the only nonvanishing light neutrino mass and $m_{h}$ the heavy mass. The other two light states stay massless. Using the single Higgs doublet of the Standard Model and calculating the loop corrections to the masses of the neutrinos does not change this qualitative picture. 

Including more Higgs doublets allows the radiative generation of a mass for one of the massless neutrinos~\cite{Grimus:1989pu}. In our numerical example we take two Higgs doublets with the Yukawa couplings, eq.(\ref{defYukawa31}), defined by the vectors $\vec{a}_{1}^{\top} = (0,0,1)$ and $\vec{a}_{2}^{\top} = (0,1,e^{i\phi})$. We ignore the effects of the Higgs sector that do not influence the neutrino masses, only the lightest neutral Higgs is required to have 125~GeV. For calculating the loop corrections we follow the example of~\cite{Grimus:2002nk}. 

\begin{figure}[htb]
\centerline{%
\begin{picture}( 350, 80)( 0, 0)
  \put( -5,10){\includegraphics[width=360pt]%
    {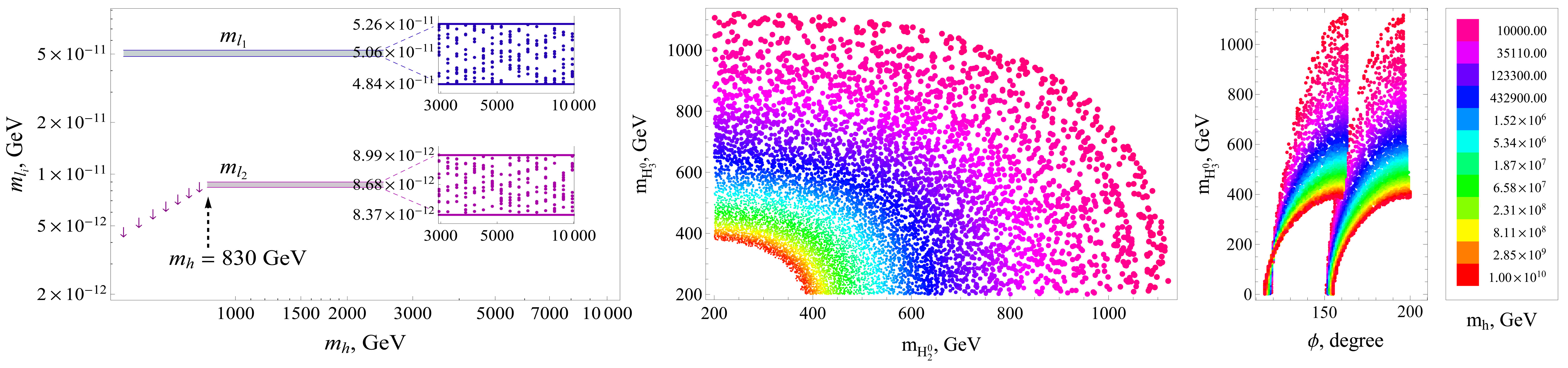}}
\put( 75,  0){{\small (a)}}
\put(205,  0){{\small (b)}}
\put(300,  0){{\small (c)}}
\end{picture}%
}
\caption{Results for the case $n_{R}=1$. The plot (a) shows the light neutrino masses in dependence on the mass of the heavy singlet. The middle (right) scatterplot (i.e. (b) and (c)) shows allowed parameter points depending on the mass of the heavy singlet $m_{h}$ through the colour code and on the masses of the heavier neutral Higgses $m_{H^{0}_{2,3}}$ (the parameter $\phi$, describing the relative phase of the Yukawa couplings).}
\label{Fig:F1H}
\end{figure}
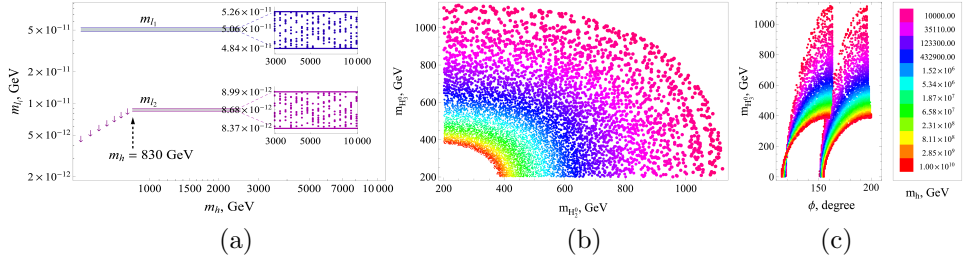

For showing the numerical result of the calculations we use the Monte Carlo method. We generate random sets of the parameters $\{ m_{h} , m_{H_{2,3}} , \phi \}$ which determine the Yukawa couplings and the size of the loop corrections. If the generated one-loop neutrino masses fulfill the measured $\Delta m^{2}_{\odot}$ and $\Delta m^{2}_{\mathrm{atm}}$ we consider the set an allowed point. As we see in Figure~\ref{Fig:F1H}(a), the variation of the heavy Higgs masses and the different Yukawa couplings allow a band of neutrino masses, that can fulfill the experimental constraint as long as the heavy singlet is heavier than 830~GeV. 

Figure~\ref{Fig:F1H}(b) shows the correlation of the masses of the heavy Higgses with the scale of the heavy singlet. The distribution of the allowed points suggest, that only the size of the Higgs mass matters, but its type, whether it is $CP$-conserving or $CP$-violating, is less important. The figure also tells us, that for a small scale of the heavy singlet, the masses of the heavy Higgses have to become very big, suggesting a decoupling limit. 

Figure~\ref{Fig:F1H}(c), finally, shows the tight correlation between the alignment of the Yukawa couplings and the required size of the Higgs masses. Only for $\cos\phi < -0.2$ we can get a rather tight prediction for the value of $\phi$ when we pick the scale of the heavy singlet and the masses of the heavy Higgses. 

\subsection{The case $n_{L}=3$ and $n_{R}=2$}
We parametrise the Yukawa couplings as 
\begin{equation} 
  \mathrm{Y}_{\! \nu}^{k} 
= \frac{\sqrt{2}}{v} 
  \left(\begin{array}{c} 
    m_{D_{2}} \vec{a}_{k}^{\top} \\ m_{D_{1}} \vec{b}_{k}^{\top} 
  \end{array}\right)
\quad\mbox{with}\enspace
m_{D_{i}}^{2} = m_{\nu_{i}} m_{h_{i}}
\enspace ,
\label{defYukawa32}
\end{equation}
where we order the masses as $m_{h_{1}} > m_{h_{2}}$ and $m_{\nu_{1}} > m_{\nu_{2}} > m_{\nu_{3}} = 0$ (at tree level). The vectors $\vec{a}_{k}$ and $\vec{b}_{k}$ describe the relative coupling stength of the Higgs doublets. At tree level we can reduce the $(3+2)\times(3+2)$ mass matrix $M_{\nu}$ according eq.(\ref{Mnu1}) to a $4\times 4$ dimensional $M_{\nu}^{\prime}$, which can be solved by the Grimus-Lavoura ansatz, with $M_{D}$ parametrized by the Casas-Ibarra ansatz, eq.(\ref{casas-ibarra}). 

Although we can get both mass differences, $\Delta m^{2}_{\odot}$ and $\Delta m^{2}_{\mathrm{atm}}$, already at tree level we perform our numerical analysis with the loop corrections for the masses of the light neutrinos included. We fix $m_{H_{1}} = 125$~GeV and $m_{h_{2}} = 100$~GeV and vary the parameters $\{ m_{h_{1}} , \vec{a}_{k}, \vec{b}_{k} , m_{H_{2}} , m_{H_{3}} \}$ with the constraint $m_{H_{2,3}} > 200$~GeV and check if the mass differences between the light neutrinos give $\Delta m^{2}_{\odot}$ and $\Delta m^{2}_{\mathrm{atm}}$. In this case, the influence of the heavier Higgses is much smaller than for the case $n_{R}=1$. This can be easily understood as the Higgs masses only influence the mass corrections to the light neutrinos. But the mass differences of the light neutrinos, $\Delta m^{2}_{\odot}$ and $\Delta m^{2}_{\mathrm{atm}}$, are already mostly determined by the tree level. 

\begin{figure}[htb]
\centerline{%
\begin{picture}( 350, 125)( 3,-5)
\includegraphics[width=12.5cm]{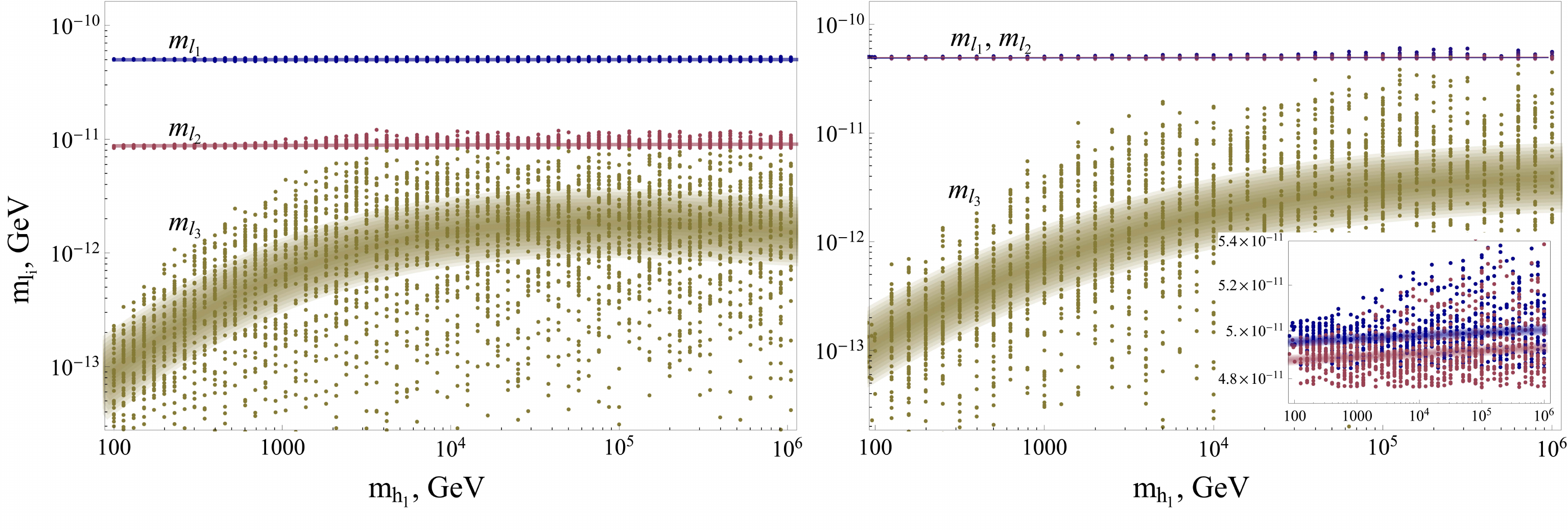}%
\put(-263, -5){{\small (a)}}
\put(- 88, -5){{\small (b)}}
\end{picture}%
}
\caption{Results for the case $n_{R}=2$, showing the light neutrino masses in dependence on the mass of the heavier singlet. The mass of the lighter heavy singlet $m_{h_{2}}$ is set to 100~GeV. Normal (inverted) neutrino mass ordering is shown on the left (right) side.}
\label{Fig:F2H}
\end{figure}

In Figures~\ref{Fig:F2H}(a) we show the solutions for the normal hierarchy of the neutrino masses. We pick a very light scale for the heavy singlet, namely just 100~GeV. An interesting observation is the reduction of the loop generated light neutrino mass with the increase of the mass of the heavier singlet beyond $10^{6}$~GeV. 
The variation of the mass of the lightest neutrino does not saturate the cosmological bound on light neutrinos of $\sum_{i} m_{\nu_{i}} < 1$~eV. 

In Figure~\ref{Fig:F2H}(b) we show the solutions for the inverted hierarchy of the neutrino masses. Again we pick a very light scale for the heavy singlet, namely just 100~GeV. In this plot the light neutrino mass increases monotonically with the mass of the heavier singlet. Even though the sum of the masses gets higher than in the normal hierarchy, we still cannot saturate the cosmological bound. So cosmology will not restrict the parameter space of our model. 

\section{Conclusions and Outlook}
Our study of the cases $n_{R} = 1$ and $n_{R} = 2$ shows, that both cases are not excluded by simple considerations of the measured neutrino mass differences. To use the data of the neutrino mixing matrix we have to assume something about the charged lepton mass matrices, which was not out goal. The case $n_{R} = 1$ predicts a tight correlation between scale of the seesaw, masses of the heavy Higgses and the values of the Yukawa coupling, suggesting a fine tuning of the Higgs sector in order to allow this scenario. Further investigation into the required Higgs sector and the allowed Yukawa couplings is neccessary to rule out this scenario. 

The case $n_{R} = 2$ still has too many free parameters to give any tight predictions. As we did not consider the charged lepton mass matrix, we could not use the neutrino mixing matrix as a constraint to our model. 

We saw in our analysis, that a treatment of the extended Higgs sector is needed. Since we are not Higgs specialists, we plan to look for a source that can easily give experimental limits on the parameters of the Higgs sector, including the Yukawa couplings. With this tool equipped we hope to support or rule out our $n_{R}=1$ scenario and to sensibly restrict the case $n_{R}=2$.

\subsubsection*{Acknowledgments}
T.G., A.J., and D.J. thank the Lithuanian Academy of Sciences for the support (the project TauPol2013, Nr.~CERN-VU-2013-3).

\end{document}